\documentclass[onecolumn,amsmath,amssymb,nofootinbib,letterpaper,11pt]{article}
\pdfoutput=1
\usepackage{jheppub}
\usepackage{ifpdf}
\usepackage[utf8]{inputenc}
\usepackage[T1]{fontenc}
\usepackage{graphicx}
\input{epsf}
\usepackage{epsfig}
\usepackage{epstopdf}
\usepackage{arcs}
\usepackage{subfigure}
\usepackage{tensor}
\usepackage{braket}
\usepackage{float}
\usepackage[dvipsnames]{xcolor}
\usepackage{amsmath}
\usepackage{morefloats}
\newcommand{\be}{\begin{equation}}
\newcommand{\ee}{\end{equation}}

\usepackage{tikz}
\usetikzlibrary{calc}   
\usetikzlibrary{topaths}
\usepackage{hyperref}
\hypersetup{
	colorlinks =WildStrawberry,
	linkcolor =red,
	pdftitle={Leading corrections to Kerr geometry},
	citecolor=NavyBlue,
	filecolor=WildStrawberry,
	urlcolor=WildStrawberry,
}

\usepackage{color}

\usepackage[toc]{appendix}
\usepackage{color,soul}
\usepackage{datetime}
\newcommand{\bea}{\begin{eqnarray}}
\newcommand{\eea}{\end{eqnarray}}
\newcommand{\ba}{\begin{eqnarray}}
\newcommand{\ea}{\end{eqnarray}}

\newcommand{\beq}{\begin{equation}}
\newcommand{\eeq}{\end{equation}}
\newcommand{\beqa}{\begin{eqnarray}}
\newcommand{\eeqa}{\end{eqnarray}}
\newcommand{\beqar}{\begin{eqnarray*}}
\newcommand{\eeqar}{\end{eqnarray*}}

\renewcommand{\c}{$c$}

\renewcommand{\href}[2]{#2}

\title{Universal criticality of thermodynamic curvatures for charged AdS black holes}
\author[a]{Seyed Ali Hosseini Mansoori,}
\author[a]{Morteza Rafiee}
\author[b]{and Shao-Wen Wei}
\affiliation[a]{Faculty of Physics, Shahrood University of Technology, P.O. Box 3619995161, Shahrood, Iran\vspace{0.1cm}}
\affiliation[b]{Institute of Theoretical and Physics Research Center of Gravitation,
	Lanzhou University,\\ Lanzhou 730000, People’s Republic of China\vspace{0.1cm}}
\vspace{0.1cm}
\emailAdd{ shosseini@shahroodut.ac.ir} 
\emailAdd{m.rafiee@shahroodut.ac.ir}
\emailAdd{weishw@lzu.edu.cn}
\abstract{In this paper, we analytically study the critical exponents and universal amplitudes of the thermodynamic curvatures such as the intrinsic and extrinsic curvature at the critical point of the small-large black hole phase transition for the charged AdS black holes. At the critical point, it is found that the normalized intrinsic curvature $R_N$ and extrinsic curvature $K_N$ has critical exponents 2 and 1, respectively. Based on them, the universal amplitudes $R_Nt^2$ and $K_Nt$ are calculated with the temperature parameter $t=T/T_c-1$ where $T_c$ the critical value of the temperature. Near the critical point, we find that the critical amplitude of $R_Nt^2$ and $K_Nt$ is $-\frac{1}{2}$ when $t\rightarrow0^+$, whereas $R_Nt^2\approx -\frac{1}{8}$ and $K_Nt\approx-\frac{1}{4}$ in the limit $t\rightarrow0^-$. These results not only hold for the four dimensional charged AdS black hole, but also for the higher dimensional cases. Therefore, such universal properties will cast new insight into the thermodynamic geometries and black hole phase transitions.}

\preprint{
}

\begin{document}
\maketitle

\section{Introduction}

Since the establishment of the four laws of black hole thermodynamics \cite{Bekenstein,Bardeen}, the study of the phase transition has been one of the increasingly active areas among the black hole physics. It also provides us an intriguing approach to peek into the microstructures of black holes. Microscopic interaction is also expected to be uncovered based on the statistical physics.

Recently, the cosmological constant was interpreted as the thermodynamic pressure and its conjugate quantity as a thermodynamic volume of a black hole.
This gives rise to new investigations of the black hole phase transition in the extended phase space \cite{Chamblin:1999tk,Chamblin:1999hg,Kubiznak}. Besides the Hawking-Page phase transition that takes place between thermal radiation and stable large black holes, the small-large black hole phase transition of a charged AdS black hole was completed, reminiscent of the liquid-gas phase transition of the van der Waals (VdW) fluid \cite{Kubiznak}. The phase transition starts at zero temperature. Then with increasing temperature, the phase transition point extends and ends at a critical point, where the first order phase transition becomes a second order one. At the critical point, it was found that it shares the same critical exponents with the VdW fluid. Subsequently, this study was generalized to other black hole backgrounds, and more phase transitions and phase structures were revealed, such as the reentrant phase transition, triple point, $\lambda$-line phase transition \cite{Kubiznak,Altamirano,AltamiranoKubiznak,Altamirano3,Dolan,Liu,Wei2,Frassino,
Cai,XuZhao,Kostouki,Hennigar,Hennigar2,Tjoa2,Ruihong,WeiSci,Sherkatghanad:2014hda}, or see \cite{Altamiranoa,Teob} for a recent review.

On the other hand, Riemannian geometry provides us with a useful tool to study some thermodynamic aspects like  phase transitions and critical behaviors \cite{reff9}. For instance, Weinhold
proposed in the thermodynamic equilibrium space a Riemannian metric defined by a Hessian of the internal energy function \cite{reff8}.
By applying the fluctuation theory to equilibrium states, Ruppeiner \cite{reff9,reff10} also introduced a different metric structure which is constructed by the second derivatives of the entropy function. It turns out that Weinhold metric is conformally related to Ruppeiner metric with the temperature $T$ as the conformal factor.

It is generally believed that a phase transition is a competition between the microscopic interaction and thermal motion of micro degrees of freedom of a black hole. Thus combining the phase transition and the thermodynamic geometry, we can investigate the microstructures of black holes. This idea was first implemented in Ref. \cite{Weiw}, where the interaction between two black hole molecules was uncovered by constructing the Ruppeiner geometry.

As suggested in Ref. \cite{reff9}, the thermodynamic scalar curvature (intrinsic curvature) scales like the correlation length of the system and goes to infinity at the critical point. As a consequence of this themodynamic curvature, interesting features are observed in some black hole systems \cite{Cai1,Cai2,Cai3,Cai4}. However, in some contradictory examples  \cite{Weiw,reff5,Sarkar:2006tg}, the scalar curvature of the Ruppeiner geometry does not diverge exactly at the critical point. To tackle this problem, a new formulation of the Ruppeiner geometry, constructed from considerations about the thermodynamic potentials that are related to the energy function (instead of the entropy) by Legendre transformations, proposed in Refs. \cite{reff12,reff13,HosseiniMansoori:2019jcs}. This new formalism of thermodynamic geometry (NTG) provides us with a one-to-one correspondence between crtical points where phase transitions occur and curvature singularities \cite{HosseiniMansoori:2019jcs}. It is worth mentioning that one of the authors of this paper obtained independently the same metric in Ref. \cite{Wei:2019uqg}.

Using such a formalism and associated features, one can find that both attractive and repulsive interactions can dominate between these micromolecules \cite{Wei:2019uqg,Wei:2019yvs}. This is quite different from the VdW fluid, where only an attractive interaction dominates. The result uncovers a significant microscopic property of charged AdS black holes. In particular, the critical behaviors were observed \cite{Wei:2019uqg,Wei:2019yvs}, based on which the correlation length can be well constructed. This study was also applied to the five dimensional neutral Gauss-Bonnet AdS black hole \cite{Weiplb}. The interesting result suggests that the interactions can keep unchanged as the system undergoes a phase transition, whereas its microstructures have a huge change. Other studies can also be found in \cite{Ghoshprd2019,XuWuYang2010,Kumara,Weipprd2020,Yerra2020,Yanga2020,Ghosh:2019pwy,HosseiniMansoori:2020yfj}.

Among the study of the phase transition, critical phenomena including critical exponents and scaling laws can reveal universal properties near a critical point for the system. So it is significant to explore critical behavior for the thermodynamic quantities. As shown in Refs. \cite{Wei:2019uqg,Wei:2019yvs}, it was found that the  intrinsic curvature goes to negative infinity at the critical point. Further, we observed a critical exponent 2 and a universal critical amplitude -1/8. These quantities disclose the critical behavior of the new formalism of the Ruppeiner geometry.

Extrinsic curvature geometry is also an interesting approach in the thermodynamic state space to explore the phase transition. It has been shown in \cite{Mansoori:2016jer} that the extrinsic curvature of a certain hypersurface in the thermodynamic space contains useful information about the location of the second-order critical point and stability of a system. It means that the extrinsic curvature shares the same divergent points and signs with specific heat on such hypersurfaces. 
This idea has also been extended to the geometrothermodynamics (GTD) geometry \cite{Zhang:2018djl}. In this paper, we aim to examine the critical exponents and universal amplitudes for the intrinsic curvature of the whole thermodynamic manifold and the extrinsic curvature on isotherm hypersurfaces immersed in the thermodynamic manifold. 

This paper is organized as follows. In the next Section, we employ the new formalism of the thermodynamic geometry (NTG) to define the thermodynamic curvatures. We then recall some basic facts about criticality behaviors of the VdW fluid in Section \ref{sec:vdw} and apply our approach to such a fluid for deriving critical exponents and amplitudes for the thermodynamic curvatures near the critical point. In Sections \ref{sec:4DB} and \ref{sec:HDB}, we examine such criticality behaviors of the thermodynamic curvatures for the AdS black holes in arbitrary dimensions. Finally, Section \ref{Conclusion} is devoted to discussions of our results.

\section{Thermodynamic curvatures}\label{sec:NTG}
In this section, we try to define the thermodynamic curvatures in simple form which can be useful for further consideration in next Sections. For this purpose, we consider the new formalism of the thermodynamic geometry (NTG) \cite{HosseiniMansoori:2019jcs} which is defined by
\begin{equation}\label{Ru1}
dl_{NTG}^2=\frac{1}{T}\left( \eta_i^{ j} \, \frac{\partial^2\Xi}{\partial X^j \partial X^l} \, d X^i d X^l \right)
\end{equation}
where $\eta_i^{ j}={\rm diag} (-1,1,...,1)$ and $\Xi$ is the thermodynamic potential and $X^{i}$ can be intensive (extensive) variables. Note that this formalism is able to explain a one-to-one correspondence between phase transitions and curvature singularities. For example, in two dimensional thermodynamic manifold, taking free energy as the thermodynamic potential, $ \Xi=F $, and $X^{i}=(T,x)$, where $x$ can be any one of the volume $V$ , charge $Q$, angular momentum $J$, or other extensive quantities, the curvature singularities correspond precisely to the phase transitions of the heat capacity at $y$ constant, $C_{y}$. For this case,
the line element is given as
\begin{equation}
dl_{NTG}^2=\frac{1}{T}\Big(-\frac{\partial^2 F}{\partial T^2} dT^2+\frac{\partial^2 F}{\partial x^2} dx^2\Big)
\end{equation}
which is in agreement with that obtained in Refs. \cite{Wei:2019uqg,Wei:2019yvs} . Using differential form of the free energy, $dF=-SdT+ydx$, one can rewrite the above metric elements as follows,
\begin{eqnarray}
dl_{NTG}^2&=&\frac{1}{T}\Big(\frac{\partial S}{\partial T}\Big)_{x}dT^2+\frac{1}{T}\Big(\frac{\partial y}{\partial x}\Big)_{T}dx^2\nonumber\\
 &=& \frac{C_{x}}{T^2} dT^2+\frac{1}{T}\Big(\frac{\partial y}{\partial x}\Big)_{T}dx^2,\label{met1}
\end{eqnarray}
in which $C_{x}=T \Big(\frac{\partial S}{\partial T}\Big)_{x}$ is the heat capacity at constant $x$.
In this paper, we are interested in applying this metric for charged AdS black holes in the extended phase space, where the cosmological constant is treated as thermodynamic pressure and its conjugate quantity as thermodynamic volume of the balck hole. Indeed, the critical behaviors of charged AdS black holes are very similar to the VdW fluid. Although, these cases have vanishing heat capacity at constant volume, i.e., $C_{V}=0$ \cite{Wei:2019yvs}, in comparison with constant heat capacity of the VdW fluid, $C_{V}=\frac{3}{2} k_{B}$, specific heat capacity of the charged AdS black hole can be treated as the limit $k_{B} \to 0^{+}$ of the VdW fluid \cite{Wei:2019yvs}. Thus, it is convenient to consider the normalized intrinsic curvature $R_{N}=C_{V}R$ which is obtained as
\begin{equation}\label{RN1}
 R_{N}=RC_{V}=\frac{(\partial_{V}P)^{2}-T^{2}(\partial_{T,V}P)^{2}+2T^{2}(\partial_{V}P)(\partial_{T,T,V}P)}{2(\partial_{V}P)^{2}}
\end{equation}
According to the equation of state for VdW-like black holes (and VdW fluid), pressure $P$ depends linearly on temperature $T$. It follows that $\partial_{T,T,V}P=0$ and therefore the normalized intrinsic curvature (\ref{RN1}) reduces to
\begin{equation}\label{RN2}
R_{N}=\frac{1}{2}\Big[1- \Big(T \frac{\partial_{V,T} P}{\partial_{V}P}\Big)^2\Big],
\end{equation}
which is used to explore the microstructures of VdW-like black holes \cite{Wei:2019yvs,Wei:2019uqg}.

On the other hand, in order to observe the VdW type criticality, one needs to keep most of thermodynamic quantities to be fixed, then the critical point depends on a fewer thermodynamic parameters. Now, it is interesting to see what happens if we fix one of  the thermodynamic variables in the thermometric manifold to form a hypersurface and investigate the critical behavior of the extrinsic curvature of such a particular hypersurface. In this regard, in Ref. \cite{Mansoori:2016jer}  it has been found that the extrinsic curvature of a hypersurface of constant extensive/intensive variable in thermodynamic geometry is divergent at the  phase transition points and it also contains useful information about stability of a thermodynamic system. Let us focus on $T=const$  isotherms in the $P-V$ diagram. Form geometrical point of view, it is equivalent to consider $T$ constant hypersufaces in the thermodynamic manifold determined by metric elements (\ref{met1}). The unit normal vector of such hypersurfaces is given by $n_{i} = (n_{T},n_{V})=(1,0)$ and $n^{i}=g^{ij}n_{j}=(\frac{\sqrt{C_{V}}}{T},0)$. Thus the extrinsic curvature can be defined by $K=\nabla_{i}n^{i}$ and finally we obtain,
\begin{equation}\label{KN1}
K_{N}=K\sqrt{C_{V}}=\frac{1}{2}\Big[1-\frac{T \partial_{V,T}P}{\partial_{V}P}\Big]
\end{equation}
In analogous with normalized intrinsic curvature, we have defined the normalized extrinsic curvature $K_{N}$. In the next sections, we investigate the behavior of these geometrical quantities near the critical point for the VdW fluid and $d$ dimensional charged AdS black holes .

\section{van der Waals fluid}\label{sec:vdw}

\begin{figure}
\centering
\includegraphics[width=7cm]{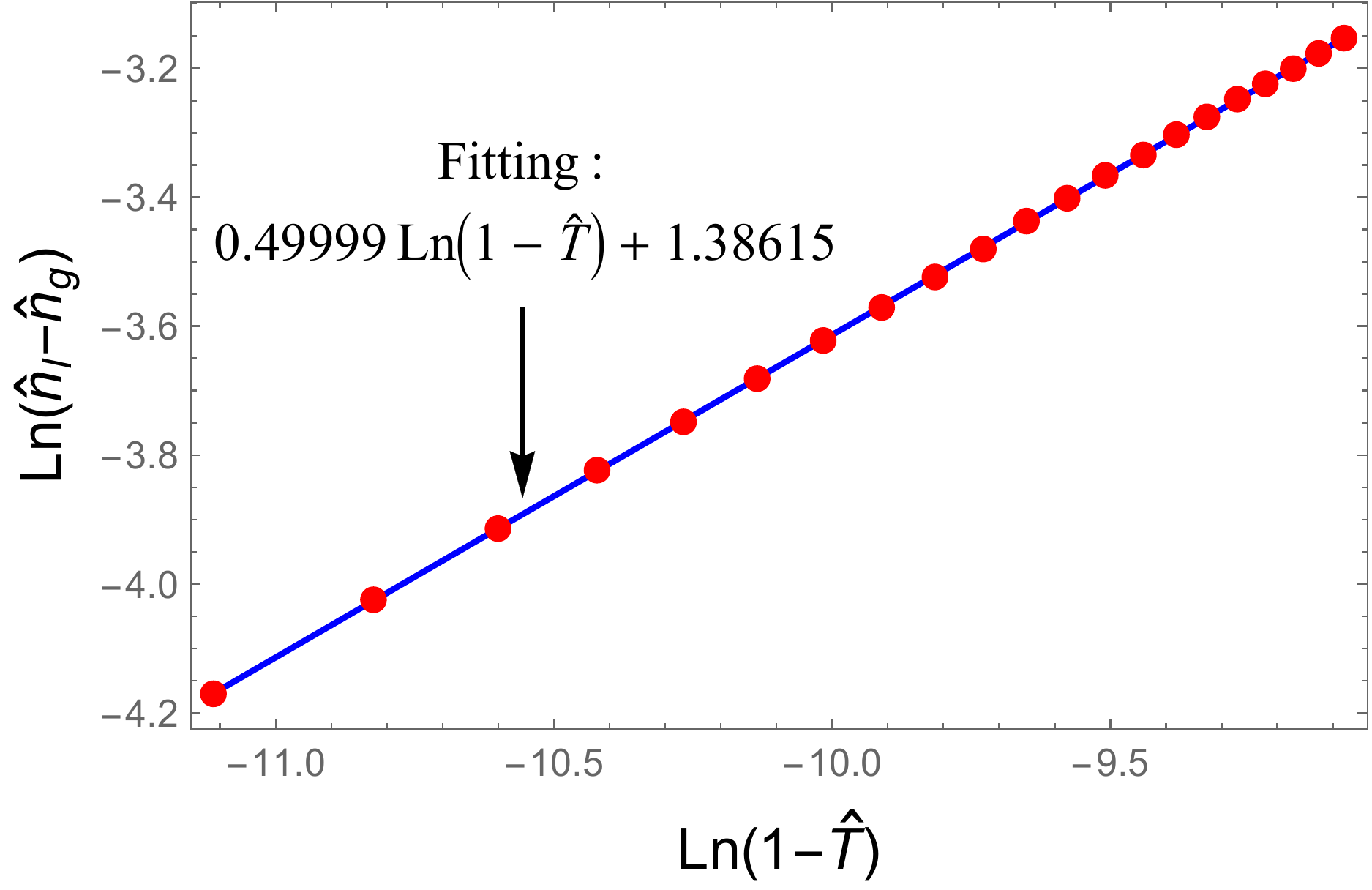}
\caption{Diagram of $\ln(n_{g}-n_{l})$ versus $\ln(1-\hat{T})$ on crossing the $\hat P$-$\hat T$ gas-liquid coexistence curve of VdW fluid. The fitted straight line for the data points (red dot) is given by $\Delta n=b (-t)^{\beta}$ with $b=3.9942$ and $\beta = 0.49999$.}\label{figdeltanvdw}
\end{figure}

\begin{figure}
\begin{center}
\subfigure[]{\label{Liquidp}
\includegraphics[width=7cm]{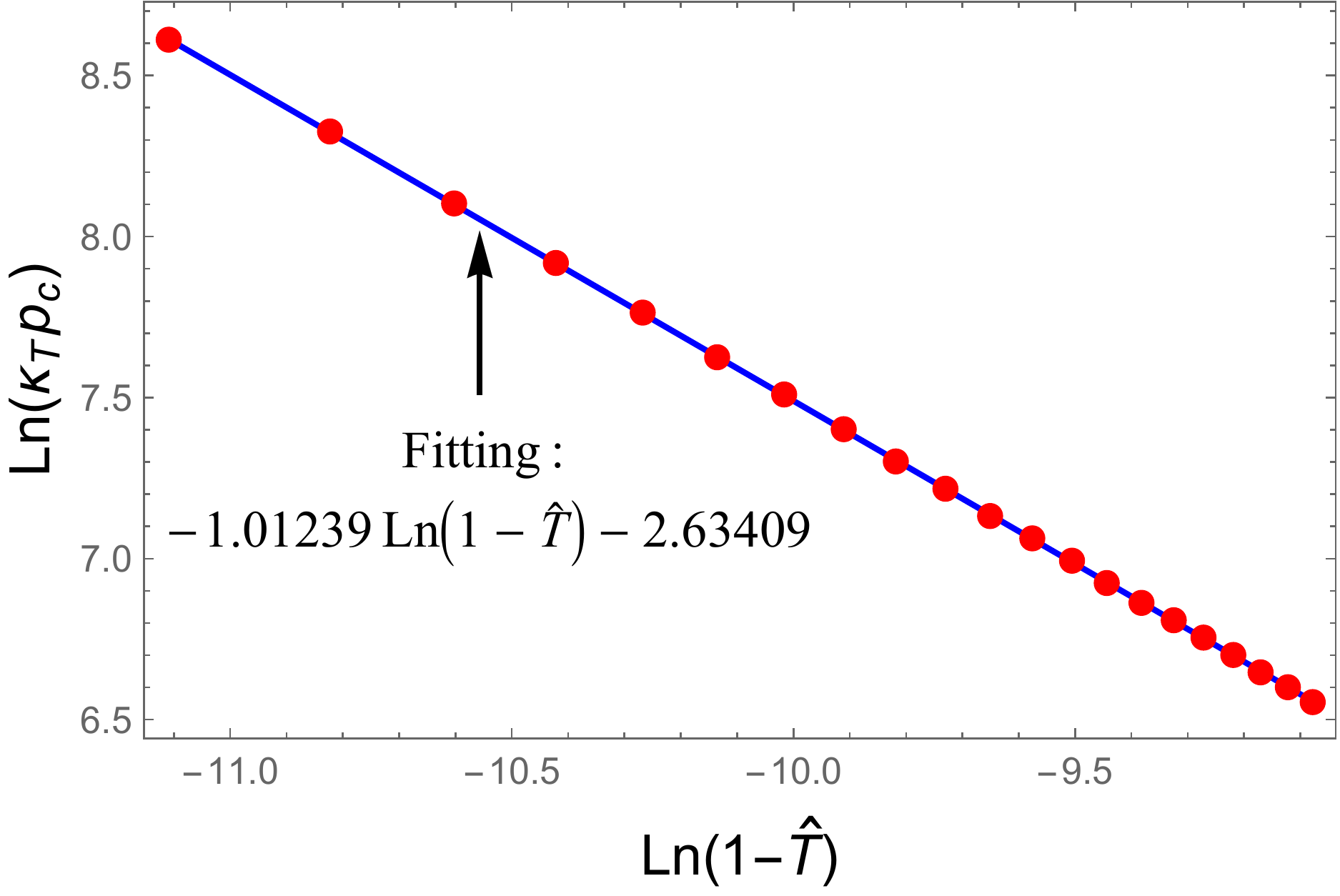}}
\subfigure[]{\label{gas}
\includegraphics[width=7cm]{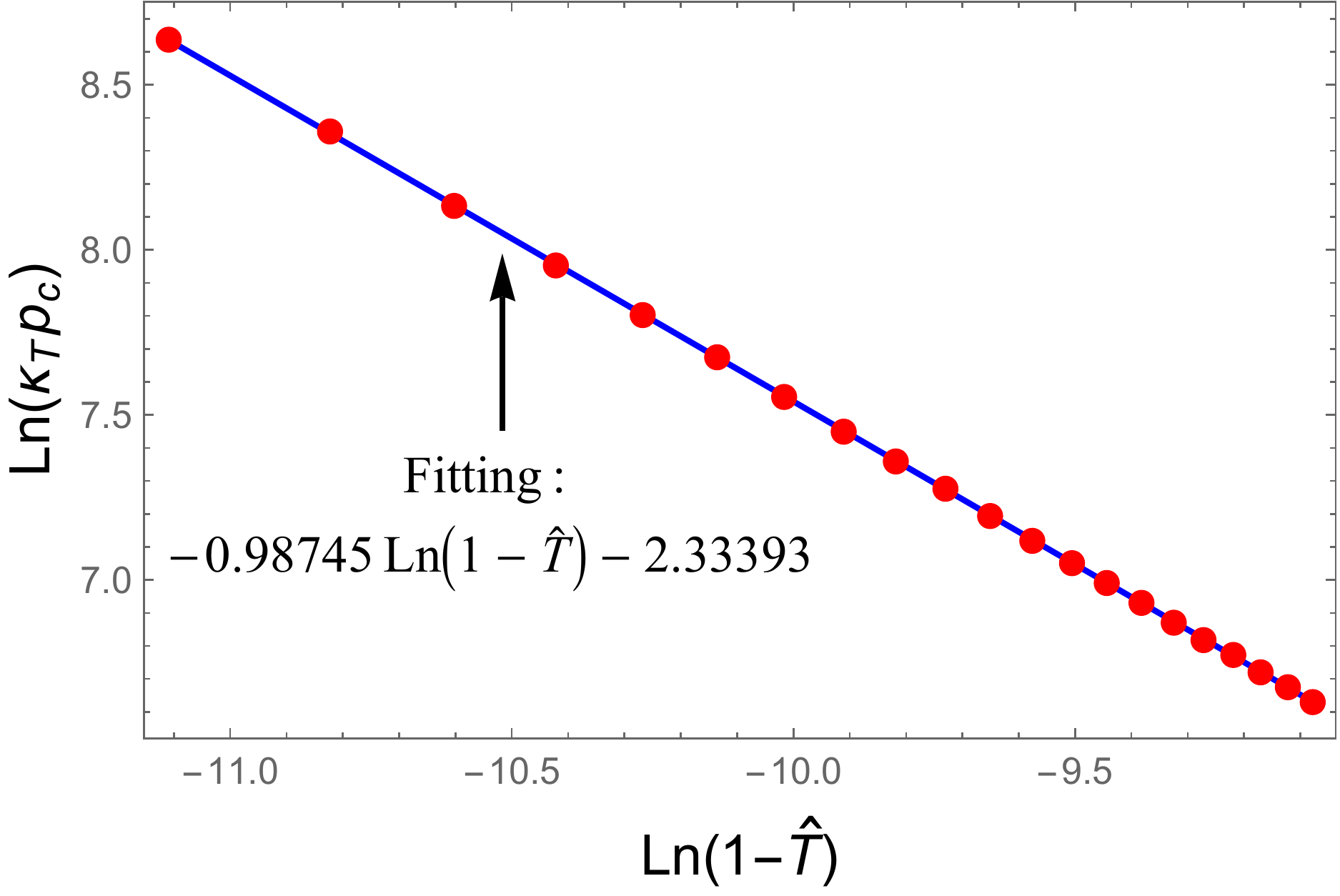}}
\end{center}
\caption{The Ln-Ln plot of $\kappa_{T}P_{c}$ versus the difference $1-\hat{T}$. The slope of the fitted blue straight line for the numerical data, described filled red circles, (a)   is -1.01239 along the coexistence saturated liquid curve (b) is -0.98745 along the coexistence saturated gas curve.}\label{fig2}
\end{figure}

\begin{figure}
\begin{center}
\subfigure[]{\label{Liquidp}
\includegraphics[width=7cm]{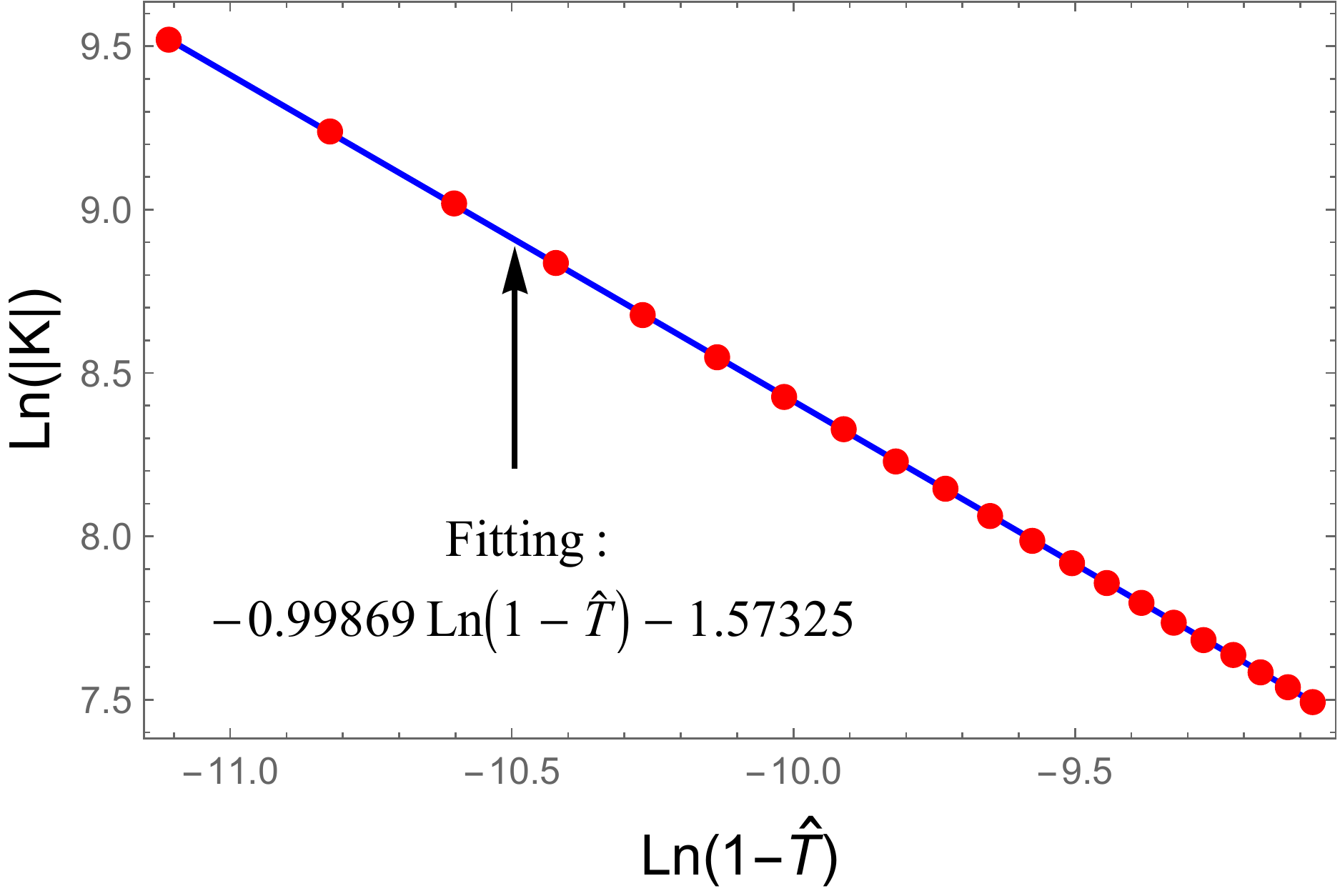}}
\subfigure[]{\label{gas}
\includegraphics[width=7cm]{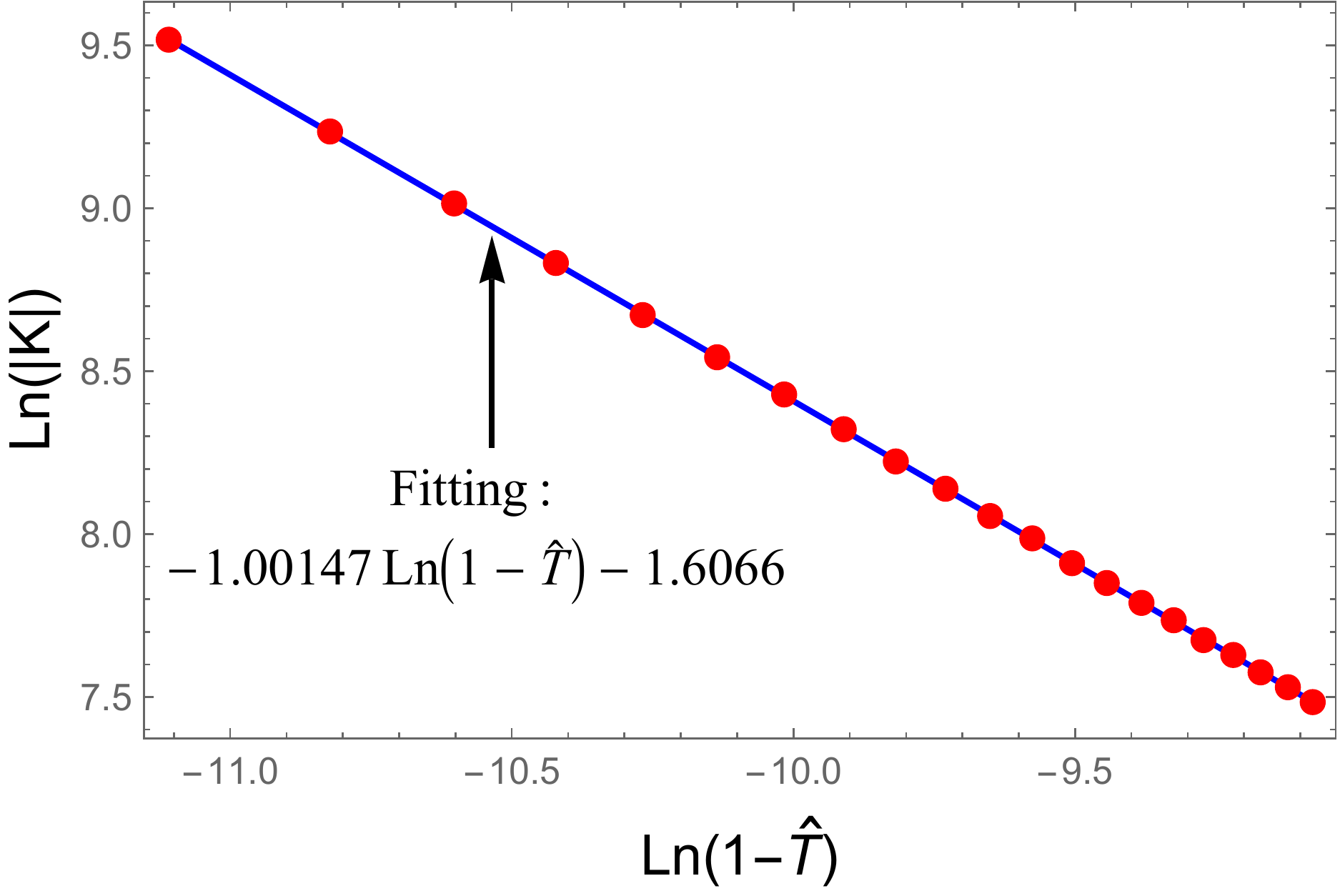}}
\end{center}
\caption{The extrinsic curvature $\ln |K|$ vs. $\ln(1-\tilde{T})$.  The numerical data are given by the red markers and the fitting formulas are determined by blue solid lines. (a) Along the coexistence saturated liquid curve, the slope is -0.99869. (b) Along the coexistence saturated gas curve, the slope is -1.00147.}\label{fig1}
\end{figure}

The first and simplest well-known example of an interacting system is the VdW fluid, which can be used to describe the first-order phase transition between gas and liquid phases. The equation of state is given by
\begin{equation}\label{EOS}
P=\frac{T}{v-b}-\frac{a}{v^2}
\end{equation}
where $v$ is the specific volume of the fluid and is related to  total volume by $v=V/N$ where $N$ the total number of all the microscopic molecules. The critical point is also determined by using the set of two conditions $(\partial_{v}P)_{T}=0$ and $(\partial_{v,v}P)_{T}=0$. By taking advantage of equation of state (\ref{EOS}), the critical point is
\begin{equation}
P_{c}=\frac{a}{27 b^2}, \quad v_{c}=3 b, \quad T_{c}=\frac{8 a}{27 b}
\end{equation}
By defining the reduced variables, $ \hat{P}=\frac{P}{P_{c}} $, $ \hat{v}=\frac{v}{v_{c}} $, and $ \hat{T}=\frac{T}{T_{c}} $, then Eq. (\ref{EOS}) becomes
\begin{equation}\label{qos1}
\hat{P}=\frac{8 \hat{T}}{3 \hat{v}-1}-\frac{3}{\hat{v}^2}
\end{equation}
which is the equation of state in the reduced parameter space \cite{Johnston}. Considering the reduced form of the number density $\hat{n} =1/\hat{v}$ between the liquid and gas phases, one can write Eq. (\ref{qos1}) in terms of the reduced fluid number density as
\begin{equation}
\hat{P}=\frac{8 \hat{T} \hat{n}}{3-\hat{n}}-3 \hat{n}^2.
\end{equation}
In the reduced parameter space, the isothermal compressibility $\kappa_{T}$ is also defined as
\begin{equation}\label{Kappa1}
 \kappa_{T}\equiv-\frac{1}{V} \Big(\frac{\partial V}{\partial P}\Big)_{T}=-\frac{1}{P_{c} \hat{v} } \Big(\frac{\partial \hat{v}}{\partial \hat{P}}\Big)_{\hat{T}}=\frac{1}{P_{c}  \hat{n}} \Big(\frac{\partial \hat{n}}{\partial \hat{P}}\Big)_{\hat{T}}
\end{equation}
Utilizing the expression for the reduced pressure in Eq. (\ref{qos1}), Eq. (\ref{Kappa1}) gives
\begin{equation}\label{kapp1}
\kappa_{T}P_{c}=\frac{(3-\hat{n})^2/6 \hat{n} }{4 \hat{T}-\hat{n}(3-\hat{n})^2}
\end{equation}
In order to expand the thermodynamic quantities around the critical point, we introduce the following new variables for simplicity
\begin{equation}\label{Eparameter}
t=\hat{T}-1, \quad \omega=\hat{v}-1, \quad \pi=\hat{p}-1.
\end{equation}
One can obtain the critical exponent $\beta$ which describes the behaviour of the liquid-gas number density difference (order parameter) along $\hat{p}-\hat{T}$ coexistence curve \cite{Johnston},
\begin{equation}\label{deltan}
\Delta n=n_{g}-n_{l} \propto (-t)^{\beta}  \quad \rm for \quad t<0.
\end{equation}
In Fig. \ref{figdeltanvdw}, we depict $\ln(n_{g}-n_{l})$ as a function of $\ln(1-\hat{T})$. The fitted straight line for the numerical data points on the lower left with $1-\hat{T}<10^{-4}$ is determined by $\Delta n=b(1-\hat{T})^{\beta}$ with $b=Exp(1.38615)=3.99942$ and $\beta=0.49999$ which are in agreement with the critical exponent $\beta=1/2$ and amplitude $b=4$ predicted in \cite{Johnston}.
On the other hand, the critical exponent $\gamma$ and $\gamma'$ determine the behavior of the isothermal compressibility $\kappa_{T}$ along the isochore $\hat{n}=1$ line and $\hat{P}-\hat{v}$ coexistence curve, receptively, as follows
\begin{equation}\label{kpc}
\kappa_{T} P_{c} \propto \bigg\{\begin{matrix}
& t^{-\gamma} & \quad \text{for} \quad t>0,\\
& (-t)^{-\gamma'} & \quad \text{for} \quad t<0.\\
\end{matrix}
\end{equation}
Therefore, by setting $\hat{n}=1$ in Eq. (\ref{kapp1}) and expanding in lower order of $t$, one can find $\gamma=1$. The ln-ln plot of the isothermal compressibility $\kappa_{T}P_{c}$ versus $(1-\hat{T})$ for $t<0$ along $\hat{P}-\hat{v}$ coexistence curve is shown in Fig \ref{fig2}. The data are seen to follow the predicted asymptotic critical behavior $\kappa_{T} P_{c} \approx (-t)^{\gamma'}$ with exponent $\gamma'=1$ in both coexistence saturated liquid and gas curves \footnote{Along $\hat{P}-\hat{v}$ coexistence curve, it is suitable to write the $\kappa_{T} P_{c}$ as a function of $\hat{P}$ and $\hat{v}$. For this purpose, we employ the alternative definition of the isothermal compressibility as follows,
\begin{equation}
\kappa_{T} P_{c}=-\frac{1}{\hat{v}} \Big(\frac{\partial \hat{v}}{\partial \hat{P}}\Big)_{\hat{T}}=-\frac{1}{\hat{v}} \frac{\{\hat{v},\hat{T}\}_{\hat{v},\hat{P}}}{\{\hat{P},\hat{T}\}_{\hat{v},\hat{P}}}=\frac{1}{\hat{v}} \Big(\frac{\partial \hat{T}}{\partial \hat{P}}\Big)_{\hat{v}} \Big(\frac{\partial \hat{T}}{\partial \hat{v}}\Big)^{-1}_{\hat{P}}.
\end{equation}
Appx. A of Ref. \cite{Mansoori:2016jer} is devoted to a brief introduction to the bracket notation.}.
Let us now examine the critical behavior of the intrinsic curvature $R_{N}$ and extrinsic curvature $K_{N}$ near the critical point. To do this, we need to rewrite Eq. (\ref{Kappa1}) as follows
\begin{equation}\label{use1}
\Big(\frac{\partial \hat{P}}{\partial \hat{n}}\Big)_{\hat{T}}=(\kappa_{T} P_{c} \hat{n})^{-1}.
\end{equation}
With the help of the lever rule for the reduced volume, the reduced density in the coexistence line can be also described by,
\begin{equation}\label{use2}
\frac{1}{\hat{n}}=\frac{\theta}{\hat{n}_{l}}+\frac{(1-\theta)}{\hat{n}_{g}},
\end{equation}
where $\theta$ is the fraction of the liquid phase, so that  for $\theta=1$, we have the pure liquid phase ($\hat{n}=\hat{n}_{l}$) and for $\theta=0$ the pure gas phase ($\hat{n}=\hat{n}_{g}$) will be dominated. Using the lever rule for the reduce number density (\ref{use2}), Eq. (\ref{use1}) can be written as
\begin{equation}
\Big(\frac{\partial \hat{P}}{\partial \hat{n}}\Big)_{\hat{T}}=\frac{1}{\kappa_{T} P_{c}}\Big(\frac{\theta}{\hat{n}_{l}}+\frac{(1-\theta)}{\hat{n}_{g}}\Big).
\end{equation}
By moving toward the critical point along the coexistence curve for $t<0$, we therefore find
\begin{eqnarray}
\Big(\frac{\partial \hat{P}}{\partial \hat{n}}\Big)_{\hat{T}}\bigg|_{n_{l}}= \frac{1}{\kappa_{T} P_{c} \hat{n}_{l}} \propto  (-t)^{\gamma'-\beta}, \\
\Big(\frac{\partial \hat{P}}{\partial \hat{n}}\Big)_{\hat{T}}\bigg|_{n_{g}}= \frac{1}{\kappa_{T} P_{c} \hat{n}_{g}} \propto (-t)^{\gamma'-\beta}.
\end{eqnarray}
As one approaches to the critical point along the isochore $\hat{v}=1$ ($\hat{n}=1$) with $t>0$, one also obtains
 \begin{equation}
 \Big(\frac{\partial \hat{P}}{\partial \hat{n}}\Big)_{\hat{T}}= \frac{1}{\kappa_{T} P_{c}} \propto  t^{\gamma}.
 \end{equation}
On the other side, the expression inside the parentheses in Eqs. (\ref{RN2}) and (\ref{KN1}) can be written as
\begin{equation}
T \frac{\partial_{V,T}P}{\partial_{V}P}=\hat{T} \frac{\partial_{\hat{V}, \hat{T}} \hat{P}}{\partial_{\hat{V}} \hat{P}}=\hat{T}\frac{\partial_{\hat{n}, \hat{T}} \hat{P}}{\partial_{\hat{n}} \hat{P}}=(1+t) \frac{\partial}{\partial t}\Big( \ln\Big(\frac{\partial \hat{P}}{\partial \hat{n}}\Big)_{t}\Big).
\end{equation}
Thus, the critical behavior of the above identity is determined by
\begin{eqnarray}
\Big(\hat{T}\frac{\partial_{\hat{n}, \hat{T}} \hat{P}}{\partial_{\hat{n}} \hat{P}}\Big)|_{n_{l},n_{g}} &\sim & \frac{(\gamma'-\beta)}{t} \quad  \text{for} \hspace{0.25cm} t<0,\\
\Big(\hat{T}\frac{\partial_{\hat{n}, \hat{T}} \hat{P}}{\partial_{\hat{n}} \hat{P}}\Big)|_{\hat{n}=1} &\sim & \frac{\gamma}{t} \hspace{1.3cm} \text{for} \hspace{0.25cm} t>0.
\end{eqnarray}
Substituting these values into Eq. (\ref{RN2}), the critical behaviors of $R_{N}$ for $t \to 0^{\pm}$ are obtained
\begin{eqnarray}\label{expaR1}
R_{N} t^{2} &\approx&- \frac{(\gamma'-\beta)^2}{2} \hspace{0.25cm} \text{for} \hspace{0.25cm} t<0,\\
R_{N} t^{2} &\approx&- \frac{\gamma^2}{2} \hspace{1.4cm} \text{for} \hspace{0.25cm} t>0.
\end{eqnarray}
In VdW fluid case with critical exponent $\beta=1/2$ and $\gamma'=\gamma=1$, we have
\begin{eqnarray}\label{inter1}
R_{N} t^{2} &\approx& - \frac{1}{8} \hspace{0.25cm} \text{for} \hspace{0.25cm} t<0,\\ \label{inter2}
R_{N} t^{2} &\approx&- \frac{1}{2} \hspace{0.25cm} \text{for} \hspace{0.25cm} t>0,
\end{eqnarray}
which exactly agrees with Refs. \cite{Wei:2019uqg,Wei:2019yvs} in $t<0$ case, whereas as $t>0$, it is consistent with Refs. \cite{Kumar:2014uba,Maity:2015rfa} (when we take $\alpha=0$ \footnote{The critical exponent $\alpha$ is for the heat capacity at constant volume.}). Note that for $t>0$ we approach to the critical point along the isochore line ($\hat{v}=1$), whereas for $t<0$ we get close to the critical point along coexistence lines between the liquid phase and the gas phase. Therefore our choice of various trajectories in $\hat{P}-\hat{v}$ diagram on approaching to the critical point leads to create a discontinuity in the value of the intrinsic curvature in Eqs. (\ref{inter1}) and (\ref{inter2}). In addition, the critical behavior of the normalized extrinsic curvature (\ref{KN1}) is given by
\begin{eqnarray}\label{expak1}
K_{N} t &\approx&- \frac{\gamma'-\beta}{2} \hspace{0.25cm} \text{for} \hspace{0.25cm} t<0\\
K_{N} t &\approx&- \frac{\gamma}{2} \hspace{1cm} \text{for} \hspace{0.25cm} t>0
\end{eqnarray}
By choosing critical exponents $\beta=1/2$ and $\gamma'=\gamma=1$ for the VdW fluid, one therefore arrives at
\begin{eqnarray}
K_{N} t &\approx& - \frac{1}{4} \hspace{0.25cm} \text{for} \hspace{0.25cm} t<0,\\
K_{N} t &\approx&- \frac{1}{2} \hspace{0.25cm} \text{for} \hspace{0.25cm} t>0,
\end{eqnarray}
 Now, we can also numerically check this critical phenomena associated with the extrinsic curvature. Near the critical point, one can numerically fit the formula by
assuming that the extrinsic curvature $K$ has the following form
\begin{eqnarray}
K\sim -(1-\tilde{T})^{-c_{K}} \quad \text{or} \quad \ln|K|=-c_{K}\ln(1-\tilde{T})+d_{K}.
\end{eqnarray}
Regarding the intercept of the straight ln-ln lines for coexistence saturated liquid and gas curves in Fig. \ref{fig1}, we find
\begin{equation}
K_{N} t= K(1-\tilde{T})\sqrt{C_{v}}=-\sqrt{\frac{3}{2}}e^{-\frac{1.57325 + 1.6066}{2}}=-0.249776\approx-\frac{1}{4} \label{oneeight}
\end{equation}
It seems that the amplitudes of the criticality of thermodynamic curvatures can be universal for any VdW-like case. Therefore, we investigate the behavior of normalized curvatures near the critical point in VdW-like black holes such as four and higher dimensional charged AdS black holes.

\section{4D charged AdS black holes}\label{sec:4DB}  

In this section, we aim to check the universality of the relations (\ref{expaR1}) and (\ref{expak1}) for a four dimensional charghed AdS black hole. The thermodynamics and phase structure of such black holes has been investigated in Ref. \cite{Kubiznak}.  The equation of state for a 4D charged AdS black hole \cite{Kubiznak} is
\begin{equation}
 P=\frac{T}{v}-\frac{1}{2\pi v^{2}}+\frac{2Q^{2}}{\pi v^{4}}, \label{stateeq}
\end{equation}
where the specific volume $v=2r_{\rm h}$ ($r_{\rm h}$ is the radius of the event horizon.).  It is interesting that a phase transition between small and large black hole phases is the similar to the liquid-gas phase transition of the VdW fluid, where the critical point is given by
$ P_{\rm c}=1/96\pi Q^{2}$, $T_{\rm c}=\sqrt{6}/18\pi Q$, and $v_{\rm c}=2\sqrt{6}Q$. In the reduced parameter space, the equation of state reads
\begin{equation}\label{RNP}
\hat{P}=\frac{8 \hat{T}}{3 \hat{v}}-\frac{2}{\hat{v}^2}+\frac{1}{3 \hat{v}^4}.
\end{equation}
Moreover, the small/large black hole coexistence curve has the analytic form  \cite{Spallucci:2013osa} as follows.
\begin{eqnarray}
 \hat{T}^{2}=\hat{P}(3-\sqrt{\hat{P}})/2,
\end{eqnarray}
where $\hat{T}=T/T_{\rm c}$ and $\hat{P}=P/P_{\rm c}$ are reduced temperature and pressure, respectively. The result has been obtained by constructing the equal area law on an isobaric curve in the $T-S$ diagram. By making use of this relation, one can obtain the reduced thermodynamic volumes for small and large black holes along the coexistence curve,
\begin{eqnarray}
 \hat{V}_{\rm s}&=&\left(\frac{\sqrt{3-\sqrt{\tilde{P}}}-\sqrt{3-3\sqrt{\hat{P}}}}{\sqrt{2\hat{P}}}\right)^{3},\label{cvs}\\
 \hat{V}_{\rm l}&=&\left(\frac{\sqrt{3-\sqrt{\hat{P}}}+\sqrt{3-3\sqrt{\hat{P}}}}{\sqrt{2\hat{P}}}\right)^{3},\label{cvl}
\end{eqnarray}
where $\hat{V}=V/V_{\rm c}$ with $V=\frac{4}{3}\pi r_{\rm h}^{3}$ and $V_{\rm c}=8\sqrt{6}\pi Q^{3}$. We now determine the critical behavior of the difference in density between the small and large black hole phases on the coexistence line. Using Eqs. (\ref{cvs}) and (\ref{cvl}), the asymptotic critical behavior of $\Delta \hat{n}$ is given by
\begin{equation}
\Delta \hat{n}=\hat{n}_{l}-\hat{n}_{s}\approx 6 \sqrt{-2t},
\end{equation}
where $n_{l}\approx 1+3 \sqrt{-2 t}$ and $n_{s} \approx1-3 \sqrt{-2 t}$ for the charged AdS black holes. Comparison of this expression with the definition in Eq. (\ref{deltan}) gives the critical exponent $\beta=1/2$. This exponent is the epitome of the mean-field theories of the second order phase transitions. Let us now consider the critical behaviors of $\kappa_{T}$ in the limit $t \to 0^{\pm}$. Differentiating the pressure (\ref{RNP}) with respect to the total volume gives
\begin{equation}
\Big(\frac{\partial \hat{P}}{\partial \hat{V}}\Big)_{\hat{T}}= -\frac{4}{9\hat{V}^{7/3}}+\frac{4}{3 \hat{V}^{5/3}}-\frac{8 \hat{T}}{9 \hat{V}^{4/3}}.
\end{equation}
By writing the above relation in terms of the expansion parameters, i.e. $t=\hat{T}-1$ and $\omega'=\hat{V}-1$ and Taylor expanding to lowest orders, we have
\begin{eqnarray}
 \Big(\frac{\partial \hat{P}}{\partial \hat{V}}\Big)_{\hat{T}}& \approx & -\frac{8}{9} t-\frac{4}{27} \omega'^2  \hspace{0.5cm} \text{for} \hspace{0.5cm} t>0, \label{pv1}\\
\Big(\frac{\partial \hat{P}}{\partial \hat{V}}\Big)_{\hat{T}}&\approx & \frac{8}{9} t-\frac{4}{27} \omega'^2 \hspace{0.8cm} \text{for} \hspace{0.5cm} t<0. \label{pv2}
\end{eqnarray}
Setting $\omega'=0$ and using Eqs. (\ref{Kappa1}) and (\ref{pv1}), we get immediately
\begin{equation}
\kappa_{T} P_{c}=\frac{9}{8}\frac{1}{t} \hspace{0.5cm} \text{for} \hspace{0.5cm} t>0.
\end{equation}
According to the definition in Eq. (\ref{kpc}), it follows that $\gamma=1$. Up to the lowest order expansion in $t$, Eqs. (\ref{cvs}) and (\ref{cvl}) give
\begin{equation}
w'^{2}_{s,l}\approx 18 t.
\end{equation}
Substituting this value into Eq. (\ref{pv2}) and using Eq.(\ref{Kappa1}), we finally arrive at
\begin{equation}
\kappa_{T} P_{c}=\frac{9}{16}\frac{1}{t} \hspace{0.5cm} \text{for} \hspace{0.5cm} t<0.
\end{equation}
Thus we can read the critical exponent $\gamma'$ from Eq. (\ref{kpc}), i.e., $\gamma'=1$. For the charged black hole cases, Eqs. (\ref{RN2}) and (\ref{KN1}) give the normalized intrinsic curvature $R_{N}$ and extrinsic curvature $K_{N}$ as
\begin{eqnarray}
R_{N}&=&\frac{(3 \hat{V}^{2/3}-1) (3 \hat V^{2/3}-1-4 \hat T \hat{V} )}{2 (1-3 \hat V^{2/3}+2 \hat T \hat V)^2},\\
K_{N}&=& \frac{1-3 \hat V^{2/3}}{2(1-3\hat V^{2/3}+2 \hat T \hat V)}.
\end{eqnarray}
Writing these expressions in terms of the expansion parameters like $t$ and $\omega'$ where $\omega'=0$ for $t>0$ and $\omega'=\omega'_{s}=-\omega'_{l}=3\sqrt{-2 t}$ for $t<0$, one can expand them up to lowest order in $t$ as follows.
\begin{eqnarray}
R_{N} t^2  \approx \bigg\{ \begin{matrix}
 -\frac{1}{2} & \text{for} & t>0,\\
 -\frac{1}{8} & \text{for} & t<0,
\end{matrix} \hspace{0.5cm} \text{and} \hspace{0.5cm} K_{N} t  \approx \bigg\{ \begin{matrix}
 -\frac{1}{2} & \text{for} & t>0,\\
 -\frac{1}{4 } & \text{for} & t<0.
\end{matrix}
\end{eqnarray}
It is interesting that above results are in consistent with those for VdW fluid. More importantly, the relations (\ref{expaR1}) and (\ref{expak1}) tend to the same universal crirical amplitudes when one chooses $\gamma=\gamma'=1$ and $\beta=1$. For $t<0$, we see that $R_{N} t^2\sim-\frac{1}{8}$, which is the same as reported in Ref. \cite{Wei:2019uqg}.

\section{Higher-dimensional charged AdS Black holes}\label{sec:HDB}

Following the study in previous sections for the VdW fluid and 4D charged AdS black holes, we here attempt to investigate the critical behavior of $R_{N}$ and $K_{N}$ for a higher dimensional charged AdS black hole and to study the effect of the dimension of the spacetime on this behavior. In the reduced parameter space, the corresponding equation of state for higher dimensional RN-AdS black holes \cite{Wei:2019yvs} becomes
\begin{equation}\label{phdimen}
 \hat{P}=\frac{\hat{V}^{-\frac{2(d-2)}{d-1}}+(d-2)\hat{V}^{\frac{2}{1-d}}\left(5-2d+4(d-3)\hat{T} \hat{V}^{\frac{1}{d-1}}\right)}{(d-3)(2d-5)}
\end{equation}
Note that there is no analytic form of the coexistence curve for small-large black holes for this case in contrast with the  four-dimensional case.  However, highly accurate fitting formulas of the coexistence curves for $d$=5-10 have been obtained in Ref. \cite{Wei:2014qwa}. Therefore, we numerically calculate the coexistence curves very near the critical point, and then obtain the critical behaviour of our interesting quantities like such as isothermal compressibility, number density, normalized intrinsic curvature, and extrinsic curvature along the coexistence curves. In order to calculate the critical exponents $\beta$, $\gamma$, and $\gamma'$, let us first to study the treatment of the isothermal compressibility $\kappa_{T}P_{c}$ and the number density difference $\Delta n$ between the small and large black holes near the critical point\footnote{Note that critical exponents $\beta$ and $\gamma$ were calculated analytically in Ref. \cite{Gunasekaran:2012dq}.}. The isothermal compressibility of the higher dimensional charged AdS black hole is calculated by using Eqs. (\ref{Kappa1}) and (\ref{phdimen}) as follows.
 \begin{equation}
\kappa_{T} P_{c}=\frac{(-3 + d) (-1 + d) (-5 + 2 d)}{2 (d-2) \left(\hat V^{-\frac{2}{d-1}} \left(2 (d-3) \hat T \hat V^{\frac{1}{d-1}}-2
   d+5\right)+\hat V^{-\frac{2 (d-2)}{d-1}}\right)}.
 \end{equation}
On approach to the critical point along the isochore $\hat{V}=1$ with $t>0$, the critical behaviour of the
isothermal compressibility $\kappa_{T}$ is given by
\begin{equation}
\kappa_{T} p_{c}\approx \frac{(d-1) (2 d-5)}{4 (d-2) t},
\end{equation}
which implies that the critical exponent $\gamma=1$. Employing Maxwell's construction, we numerically calculate the behaviour of $\Delta n$ and $\kappa_{T}P_{c}$ with $t<0$ for $d$=5-10 along $\hat{P}-\hat{T}$ coexistence curve and $\hat{P}-\hat{V}$ coexistence curve, respectively, in Table \ref{tab1}.

\begin{table}[h]
\begin{center}
\begin{tabular}{c ||c||rrrrrr}
  \hline\hline
      Quantity & Coefficient   &   $d$=5 & $d$=6 & $d$=7 & $d$=8 & $d$=9 & $d$=10 \\\hline
 & $c_{n}$  &  0.50004 & 0.50004 & 0.50005 & 0.50005 & 0.50006 & 0.50006 \\[-1ex]
\raisebox{1.5ex}{$\ln (\Delta n)$}
 & $d_{n}$ & 2.17107 & 2.22602 & 2.28272 & 2.33658 & 2.38664 & 2.43293 \\\hline
 & $\c_{\kappa}$ & 1.00901 & 1.00938 & 1.00982 & 1.01029 & 1.01076 & 1.01122 \\[-1ex]
\raisebox{1.5ex}{$\ln |\kappa_{T}P_{c}|$ (LBH)}&
 -$d_{\kappa}$ & 0.29237 & 0.02490 & 0.18017 & 0.34698 & 0.48771 & 0.60945\\\hline
 & $c_{\kappa}$  & 0.99080 & 0.99044 & 0.98999 & 0.98952 & 0.98904 & 0.98857 \\[-1ex]
\raisebox{1.5ex}{$\ln |\kappa_{T}P_{c}|$ (SBH)}
 & -$d_{\kappa}$ & 0.07027 & 0.20613 & 0.42211 & 0.60038 & 0.75259 & 0.88563 \\ \hline\hline
\end{tabular}
\caption{Fitting values of the slope and intercept of $\ln \Delta n=c_{n} \ln(1-\hat{T})+d_{n}$ and $\ln|\kappa_{T}P_{c}|=-c_{\kappa}\ln(1-\hat{T})-d_{\kappa}$ straight line for
  coexistence saturated small-large black hole curves.}\label{tab1}
\end{center}
\end{table}

According to the data appeared in Table. \ref{tab1}, it is obvious that the values of the critical exponent $\beta=c_{n}$ and $\gamma'$, which is the average of two slopes $c_{\kappa} (\rm SBH)$ and $c_{\kappa} (\rm LBH)$, are not deeply affected by increasing dimension of the spacetime, i.e.,
\begin{equation}
\beta \approx \frac{1}{2}, \hspace{1cm} \gamma'\approx 1
\end{equation}
which are the same as those for VdW fluid and 4D charged AdS black holes. Therefore, one can expect that, independent of changing number of the dimension, the normalized intrinsic curvature $R_{N}$ and extrinsic curvature $K_{N}$ show the universal critical behavior around the critical point. Utilizing Eqs. (\ref{RN2}), (\ref{KN1}), and (\ref{phdimen}), one can easily find the criticality of thermodynamic curvatures for $t>0$ are given by
\begin{eqnarray}
R_{N} t^2 \approx -\frac{1}{2}, \quad K_{N}t\approx -\frac{1}{2},
\end{eqnarray}
which indicate a universal behaviour of thermodynamics curvatures near the critical point similar to VdW fluid and 4D charged AdS black holes, i.e., they do not depend on the details of the physical system. Let us now examine the universality by approaching significantly closer to the critical point along small/large black holes coexistence line. The coefficients of the numerical fitting of the ln-ln formulas of the $R_{N}$ and $K_{N}$ versus $(1-\hat{T})$ have been collected in Table \ref{tab2}.

\begin{table}[h]
\begin{center}
\begin{tabular}{c ||c||rrrrrr}
  \hline\hline
      Quantity & Coefficient   &   $d$=5 & $d$=6 & $d$=7 & $d$=8 & $d$=9 & $d$=10 \\\hline
 & $c_{R}$  &  2.01116 & 2.01297 & 2.01455 & 2.01598 & 2.01729 & 2.01851 \\[-1ex]
\raisebox{1.5ex}{$\ln |R_{N}|$ (SBH)}
 & -$d_{R}$ & 2.21562 & 2.23775 & 2.25709 & 2.27454 & 2.29054 & 2.30541 \\\hline
 & $c_{R}$ & 1.98875 & 1.98693 & 1.98532 & 1.98388 & 1.98255 & 1.98132 \\[-1ex]
\raisebox{1.5ex}{$\ln |R_{N}|$ (LBH)}&
 -$d_{R}$ & 1.94227 & 1.92008 & 1.90054 & 1.88292 & 1.86674 & 1.85169\\\hline
 & $c_{K}$  & 1.00567 & 1.00657 & 1.00737 & 1.00808 & 1.00874 & 1.00935 \\[-1ex]
\raisebox{1.5ex}{$\ln |K_{N}|$ (SBH)}
 & -$d_{K}$ & 1.45537 & 1.46644 & 1.47611 & 1.48484 & 1.49284 & 1.50028 \\\hline
 & $c_{K}$ & 0.99446 & 0.99355 & 0.99275 & 0.99202 & 0.99136 & 0.99074 \\[-1ex]
\raisebox{1.5ex}{$\ln |K_{N}|$ (LBH)}&
 -$d_{K}$ & 1.31866 & 1.30756 & 1.29779 & 1.28898 & 1.28089 & 1.27336 \\ \hline\hline
\end{tabular}
\caption{The coefficients of the numerical fitting of the ln-ln formulas for coexistence saturated small black holes (SBH) and coexistence saturated large black holes (LBH).}\label{tab2}
\end{center}
\end{table}

Taking numerical error into account, the slopes indicate that the critical exponent for $R_{N}$ and $K_{N}$ must be $c_{R}=2$ and $c_{K}=1$, which is consistent with those of VdW fluid. Considering the interception of the straight ln-ln line (i.e., $d_{R}$ and $d_{K}$), the amplitude associated with these critical exponents are given in Table \ref{tab3}.

\begin{table}[h]
\begin{center}
\begin{tabular}{c ||rrrrrr}
  \hline\hline
      Quantity    &   $d$=5 & $d$=6 & $d$=7 & $d$=8 & $d$=9 & $d$=10  \\\hline
 $R_{N} t^2$  & -0.12506 & -0.12507 & -0.12508 & -0.12509 & -0.12510 & -0.12511 \\\hline
 $-\frac{(\gamma'-\beta)^2}{2}$  & -0.12493 & -0.12493 & -0.12493 & -0.12493 & -0.12492 & -0.12492\\\hline
 $K_{N} t$  & -0.24982 & -0.24982 & -0.24984 & -0.24985 & -0.24986 & -0.24987\\\hline
 $-\frac{(\gamma'-\beta)}{2}$  & -0.24993 & -0.24993 & -0.24993 & -0.24993 & -0.24992 & -0.24992\\ \hline\hline
\end{tabular}
\caption{Critical amplitudes of thermodyanic curvatures near the critical point.}\label{tab3}
\end{center}
\end{table}
Clearly, the change of amplitudes is negligible by increasing the number of spacetime dimensions. Therefore, we can conclude that the critical behavior of the normalized thermodynamic curvatures for $t<0$ along the coexistence curve are \begin{equation}
R_{N} t^2 \approx -\frac{1}{8}, \hspace{1cm} K_{N} t \approx -\frac{1}{4}.
\end{equation}
which show universal behavior independently of the number of spacetime dimensions. Moreover, they confirm universal amplitudes obtained in Eqs. (\ref{expaR1}) and (\ref{expak1}). Note that in \cite{Wei:2019uqg,Wei:2019yvs}, it is shown that the values of $R_{N} t^2$ depends on the spacetime number $d$, which is different from that observed here. The reason is that the calculation accuracy in \cite{Wei:2019uqg,Wei:2019yvs} is lower, and in this paper we only keep the lowest order of the expansion.

\section{Conclusions and discussions}
\label{Conclusion}

In this paper, we studied the critical behaviors of the normalized intrinsic curvature $R_N$ and the extrinsic curvature $K_N$. The universal properties are analytically checked when the thermodynamic quantities are expanded to the lowest order.

First, we dealt with the VdW fluid. Employing the equation of state, we found that $R_N$ has critical exponent 2 and $K_N$ has critical exponent 1 when the temperature parameter $t$ approaches zero from two sides. The result of $R_N$ is also consistent with that of \cite{Wei:2019uqg,Wei:2019yvs}. By making use of this result, we can construct the universal constants $R_Nt^2$ and $K_Nt$. Adopting the numerical calculation, we find that $R_Nt^2=-\frac{1}{8}$ for $t\rightarrow0^-$ and $-\frac{1}{2}$ for $t\rightarrow0^+$, whereas $K_Nt=-\frac{1}{4}$ for $t\rightarrow0^-$ and $-\frac{1}{2}$ for $t\rightarrow0^+$.

Next, we applied the treatment to charged AdS black holes. For the four dimensional charged AdS black hole, $R_N$ and $K_N$, respectively, share the same critical exponents. Universal amplitude $R_Nt^2=-\frac{1}{2}$ for $t\rightarrow0^+$ and $-\frac{1}{8}$ for $t\rightarrow0^-$  and also, $K_Nt=-\frac{1}{2}$ and $-\frac{1}{4}$ for $t\rightarrow0^+$ and $t\rightarrow0^-$, respectively.

We also generalized the result to higher spacetime dimensions. When the temperature tends to the critical value, the change of the number density $\Delta n$ before and after the small and large black holes has critical exponent $\frac{1}{2}$, which indicates it can act as an order parameter to describe the small and large black hole phase transition. $\kappa_TP_c$ also presents a critical exponent 1 near the critical point when the black hole system approach the critical point along the coexistence small and large black hole curves. Moreover, when $t\rightarrow0^+$, we found $R_{N} t^2=-\frac{1}{2}$ and $K_{N}t=-\frac{1}{2}$. And when $t\rightarrow0^-$, $R_{N} t^2=-\frac{1}{8}$ and $K_{N}t=-\frac{1}{4}$. Significantly, these critical amplitudes are independent of the spacetime number $d$. The numerical calculations also confirm this results.

In summary, in this paper we have investigated the critical behaviors of the normalized intrinsic and extrinsic curvature of the NTG geometry. Universal amplitudes at the critical point are also calculated. These will uncover the underlying universal critical properties of the charged AdS black holes. The generalization to other AdS black hole backgrounds are also expected.

\section*{Acknowledgements}
We are grateful to Yu-Xiao Liu and R. B. Mann for their careful reading and extremely helpful discussions and comments on this work. Shao-Wen Wei was supported by the National Natural Science Foundation of China (Grant No. 11675064) and the Fundamental Research Funds for the Central Universities (Grants No. lzujbky-2019-it21).

\end{document}